\documentclass[12pt]{amsart}
\usepackage[left=3cm,top=3cm,right=3cm,bottom=3.5cm]{geometry}
\usepackage[utf8]{inputenc}
\usepackage{import} 
\usepackage{comment}
\usepackage{fontenc}
\usepackage{subcaption}
\usepackage{hyperref}
\usepackage{amsmath} 
\usepackage{amsthm}
\usepackage{amssymb}
\usepackage{graphicx} 
\usepackage{tikz}  
\usepackage[labelfont=it, textfont=it]{caption}
\usepackage{pdfpages}
\usepackage{mathrsfs} 
\usepackage[makeroom]{cancel}
\usepackage{bm} 
\usepackage{yhmath}
\usepackage{setspace}
\usepackage{enumerate}

\newtheorem{theorem}{Theorem}

\newtheorem{definition}{Definition}

\newtheorem{lemma}[theorem]{Lemma}
\newenvironment{Proof}[1][Proof]
  {\proof[#1]\leftskip=0.2cm\rightskip=0.2cm}
  {\endproof}
\theoremstyle{definition}
\newtheorem{remark}{Remark}

\newcommand{\al}{\alpha}
\newcommand{\be}{\beta}
\newcommand{\ga}{\gamma}
\newcommand{\de}{\delta}
\newcommand{\De}{\Delta}
\newcommand{\ep}{\epsilon}
\newcommand{\si}{\sigma}

\newcommand{\ta}{\theta}
\newcommand{\la}{\lambda}
\newcommand{\La}{\Lambda}

\newcommand{\ud}{\mathrm{d}}

\DeclareMathOperator*{\argmin}{arg\,min}

\newcommand{\artanh}{\operatorname{artanh}}

\newcommand{\T}{^{\operatorname{T}}}
\newcommand{\D}{^{\dagger}}
\newcommand{\tr}{\operatorname{tr}}

\newcommand{\mc}[1]{\mathcal{#1}}

\title{Extremal States of Multipartite Quantum Spin Systems}
\author{Oskar Olander}

\begin{document}
\setlength{\jot}{15pt} 
\onehalfspacing

\maketitle

{\small
\section*{Abstract}
\addcontentsline{toc}{section}{Abstract}
We consider two quantum spin models on $k$-partite graphs. The system is assumed to be invariant under all permutations within any of the $k$ sets. In the limit of infinitely many particles, the Gibbs state is described as a mixture of independent spins due to a version of the Quantum de Finetti theorem. 

First, we study a spin-1/2
antiferromagnetic Heisenberg model and identify its three phases. At low temperature all $k$ sets are magnetised but cancel each other, at medium temperature there is a net magnetic field and at high temperature there is no magnetisation. The magnetisation temperature is a solution to a $k$-degree polynomial.

Second, we allow a general spin $s$ and identify all orthogonally invariant models which have frustration. The criterium for a frustration-free model is that the $k$-sets can be partitioned into two parts, where the only antiferromagnetic interaction is between the two parts.}

\section{Introduction}
\noindent
In this paper we study models of magnetisation on complete $k$-partite graphs with permutation symmetry. These models exhibits interesting behaviours such as frustration, where the energy cannot be minimized simultaneously among all interacting pairs. Complete $k$-partite graphs provide a natural setting in which these effects appear, while remaining sufficiently simple to allow for a detailed analysis.

Our main tool is a $k$-partite extension of the Quantum de Finetti theorem.  The Quantum de Finetti theorem identifies the extremal exchangeable states as product states, determined by a single density matrix replicated at every site. The relevance of this result for models of magnetisation on the complete graph was first recognized by Fannes, Spohn, and Verbeure, who used it to derive the mean-field equation, a nonlinear self-consistency relation for the one-site density matrix of an extremal state \cite{Fannes}. In the $k$-partite setting the picture is very similar, except that one must determine $k$ density matrices rather than a single one.

We apply this framework to two different quantum spin systems. The first is the spin-1/2 antiferromagnetic Heisenberg model. The second is an O($2s+1$)-invariant generalization of the Heisenberg model to higher spin. Related models in the bipartite case were previously studied by Björnberg, Ryan and Rosengren using representation-theoretic methods rather than de Finetti techniques \cite{Ryan}.

For the spin-1/2 Heisenberg model, we describe the Gibbs states and phase transitions, see Theorem \ref{thm:phase}. And for the O($2s+1$)-invariant model, we identify all such models which have frustration for large $\be$, see Theorem \ref{thm:graph}.

Our work is also related to several recent developments. In \cite{Warzel}, Manai and Warzel analyse permutation invariant spin systems using Berezin-Lieb inequality and the Laplace principle. This is analogous to using Quantum de Finetti and the variational principle. In \cite{GD}, Athanasopoulos and Ueltschi analyse the classical Ising model on the triangular lattice, which exhibits frustration for certain choices of parameters. More generally, there has recently been renewed interest in magnetisation models on complete graphs, including $q$-deformations of the Heisenberg model; see, for example, Ballesteros et al \cite{Shore}. Although we assume a quite strict permutation invariance, the idea of approximating Gibbs states as product states is more general and can be used in a wider range of models, see for instance \cite{Bergamaschi2022}.




\subsection{Models}
The models have a set $\Lambda$ of $n$ identical spins that are partitioned into $k$ subsets $\La_i$, which we will refer to as \textit{needles}. 
$$ \Lambda=\Lambda_1 \sqcup 
\La_2\sqcup \dots\sqcup \La_k $$
The reason for calling the subsets needles is that the models behave quite similarly to a compass with $k$ needles. Denote the size of the needles by $|\La_i|=n_i$. For convenience, label the spins in a way such that $i\in \Lambda_j$ if and only if $i\equiv j$ modulo $k$. This means that the index $i$ labels both a spin and a needle. In particular, two spins $i,j$ are in the same needle if and only if $i\equiv j$. Let $\Phi_{ij}$ denote the interaction between spin $i$ and $j$. It is a Hermitian operator acting on $\mathbb{C}^{2s+1}\otimes \mathbb{C}^{2s+1}$, where $s$ denotes the spin number. Assume $\Phi_{i'j}=\Phi_{ij}$ if $i\equiv i'$. In order to tame the infinite range interaction, we normalise the Hamiltonian with $n$: 
\begin{equation}\label{eq:Ham}
    H_n=\frac{1}{n}\sum_{i,j\in \La} \Phi_{ij}
\end{equation}
Firstly we consider the antiferromagnetic Heisenberg model for spin $1/2$. Here we assume that the needles have no internal interaction:
\begin{equation}\label{eq:H1}
    H_{n}^{\mathrm{AF}}=+\frac{1}{n}\sum_{i\not \equiv j} \vec{S}_i\cdot \vec{S}_j
\end{equation}
where $\vec{S}_i=(S_i^1,S_i^2,S_i^3)$ is the spin operator acting on spin $i$. Spin operators are defined by the Lie algebra $\mathfrak{su}(2)$ given by $[S^a_i,S^b_j]=i\de_{ij}\ep_{abc} S^c_i$ where $a,b,c=1,2,3$, and the representation labelled by $s$ given by $\vec{S}_i\cdot \vec{S}_i=s(s+1)\bm{1}$.

The second model is the most general O$(2s+1)$ invariant two body interaction, without internal interaction in the needles:
\begin{equation}\label{eq:H2}
    H_{n}^{\mathrm{O}}=\frac{1}{n}\sum_{i\not \equiv j} (u_{ij} T_{ij} + v_{ij}Q_{ij})
\end{equation}
where the operators $T_{ij}$ and $Q_{ij}$ act on site $i$ and $j$ and are defined by:
$$ Q_{ij} =\sum_{\al,\be=-s}^s|\al\al\rangle \langle \be\be|,\quad T_{ij}=\sum_{\al,\be=-s}^s|\al\be\rangle \langle \be\al|, $$ 
where $|\al\rangle $ is an eigenvector to $S_i^3$ with eigenvalue $\al\in\{-s,-s+1,\dots,s\}$. The real numbers $u_{ij}$ and $v_{ij}$ satisfy $u_{ij}=u_{i'j'}$ and $v_{ij}=v_{i'j'}$ if $i\equiv i'$ and $j\equiv j'$. We allow $u_{ij},v_{ij}$ to be zero, however we assume that all spins are connected, meaning that there is no partition of $\Lambda=\La'\sqcup \La''$ such that there is no interaction between $\La'$ and $\La''$.
\subsection{Definitions} The state of our system is a product of $k$ Hilbert spaces corresponding to each needle. These are in turn a power of $n_i$ spins:
\begin{equation}\label{la:Hilbert}
\mathcal{H}_1^{\otimes n_1}\otimes \mathcal{H}_2^{\otimes n_2}\otimes \dots \otimes \mathcal{H}_k^{\otimes n_k},
\end{equation}
where each $\mathcal{H}_i\cong \mathbb{C}^{2s+1}$. We assume for every $i$ that $n_i=n_i(n)$ is an increasing function of $n$ such that the following limits exists and are positive: 
\begin{equation}
a_i=\lim_{n\rightarrow\infty} \frac{n_i}{n}
\end{equation} 
We refer to the $a_i\in (0,1]$ as \textit{relative sizes}. 

We will start our analysis with any sequence of density matrices  $\rho=\{\rho_n\}_{n=1}^\infty$ acting on (\ref{la:Hilbert}) such that $\rho_{n}=\tr_{n'-n}\rho_{n'}$ for all $n'\geq n$. So far there are no additional constraints other than the partial trace condition. The Hamiltonian in equation (\ref{eq:Ham}) is invariant under the following group of permutations:
\begin{equation}
    \mathrm{S}_{n_1}\times \mathrm{S}_{n_2}\times \dots \times \mathrm{S}_{n_k}
\end{equation} 
where $\mathrm{S}_{n_i}$ is the symmetric group acting on $\mathcal{H}_i^{\otimes n_i}$ in the natural way.  We will be interested in density matrices that share this symmetry.

\begin{definition}[$k$-exchangeable] Let $\rho=\{\rho_{n}\}_{n\geq 1}$ be a sequence of density matrices acting on (\ref{la:Hilbert}), such that $\rho_n=\tr_{n'-n} \rho_{n'}$ for all $n'\geq n$. We say that $\rho$ is $k$-exchangeable if $\rho_n$ is invariant under the permutations $\mathrm{S}_{n_1}\times \mathrm{S}_{n_2}\times \dots\times \mathrm{S}_{n_k}$, for all $n$.
\end{definition}
\noindent 
Any 1-exchangeable state is a mixture of systems of independent spins. This is the famous quantum de Finetti Theorem. We will formulate it for our $k$-partite setting. This version of de Finetti is well-known, but for completeness we outline the proof in Appendix \ref{appendix}.

\begin{theorem}[$k$-partite quantum de Finetti]\label{thm:definetti} Let $\{\rho_{n}\}_{n\geq 1}$ be a $k$-exchangeable sequence of density matrices on the Hilbert space $\mathcal{H}_1^{\otimes n_1}\otimes \mathcal{H}_2^{\otimes n_2}\otimes \dots \otimes \mathcal{H}_k^{\otimes n_k}$. Then there exists a unique probability measure $\mu$ on vectors of density operators $(\tau_1,\tau_2,\dots,\tau_k)$, where $\tau_i$ is an operator on $\mathcal{H}_i$, such that for all $n$:
\begin{equation}\label{eq:deFinetti}
    \rho_{n}=\int \tau_1^{\otimes n_1}\otimes \tau_2^{\otimes n_2}\otimes\dots\otimes  \tau_k^{\otimes n_k}\ud \mu(\tau_1,\tau_2,\dots,\tau_k)
\end{equation}
\end{theorem}
\noindent Using Theorem \ref{thm:definetti} we can write any extremal $k$-exchangeable state by a vector of density matrices $\tau=(\tau_1,\dots,\tau_k)$ as:
\begin{equation}\label{eq:state}
\rho_n=\bigotimes_{i=1}^k\tau_i^{\otimes n_i} 
\end{equation}
Given a vector $\tau=(\tau_1,\dots,\tau_k)$ the specific free energy for our model (\ref{eq:Ham}) is:
\begin{align}\label{eq:free}
f_\be(\tau)&=\lim_{n\rightarrow\infty}\frac{1}{n}\big( \tr(\rho_n H_n)+\frac{1}{\be}\tr(\rho_n\ln\rho_n)\big)= \nonumber \\
       & =\sum_{i,j=1}^ka_ia_j\tr(\tau_i\otimes \tau_j \Phi_{ij})+\frac{1}{\beta}\sum_{i=1}^k a_i\tr(\tau_i\ln \tau_i)
\end{align}

\begin{definition}[Extremal $k$-exchangeable Gibbs state]
An extremal $k$-exchangeable Gibbs state is a vector of density matrices $\tau=(\tau_1,\dots,\tau_k)$ which minimises $f_\be(\tau)$ given by (\ref{eq:free}). Denote the set of such states by $\mathcal{G}_{k}$.
\end{definition}
\noindent Specifically for spin 1/2, we can parametrise $\tau_i$ as:
\begin{equation}\label{eq:tau}
    \tau_i=\frac{I}{2}+\frac{\vec{y}_i}{a_i}\cdot \vec{S}_i
\end{equation}
where $I$ is the identity matrix. The vectors $\vec{y}_i\in \mathbb{R}^3$ have magnitude $y_i=|\vec{y_i}|\leq a_i$. We refer to the $\vec{y}_i$ as \textit{magnetisation vectors}. They are the average spin of a needle. Define a difference $\Delta y :=y_1-y_2-\dots-y_k$. The sign of $\Delta y$ will determine the phase of model (\ref{eq:H1}). Also let $x_i=\frac{y_i}{a_i}$ denote the normalised magnetisation.

\begin{definition}[Frustration]
If there exists an extremal $k$-exchangeable Gibbs state $\tau=(\tau_1,\dots,\tau_k)$, a pair $i,j\in \{1,\dots,k\}$ and two unitary matrices $U,U'$ such that:   
    $$ \tr( \tau_i\otimes  \tau_j\Phi_{ij})> \tr(U^\dagger \tau_iU\otimes {U'}^\dagger \tau_jU'\Phi_{ij})  $$
we say that the model (\ref{eq:Ham}) has frustration.
\end{definition}
  \noindent Frustration means that it is possible decrease the energy from at least one interaction-term while keeping the entropy constant. Frustration can only occur when $\be$ is sufficiently large. We want to find all models which are frustration free for all $\be$.

  \subsection{Main results}
\begin{figure}[h]
    \centering
    \begin{tikzpicture}
    \node[anchor=south west, inner sep=0] (cat) at (0,0)
        {\includegraphics[width=0.6\linewidth]{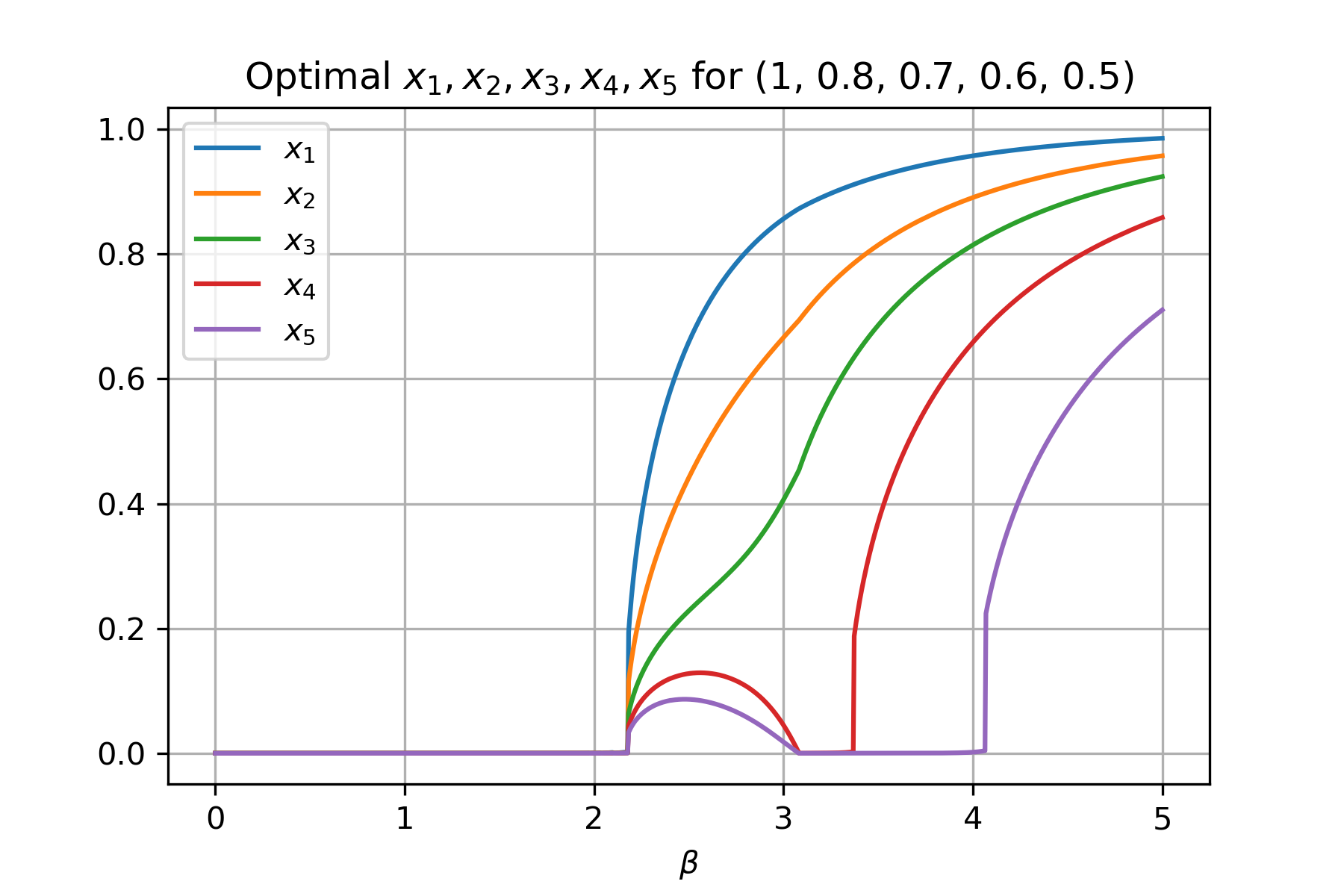}};

    \begin{scope}[x={(cat.south east)}, y={(cat.north west)}]

        \node   at (0.45,0.2) {\tiny $\beta_1$};
        \node  at (0.59,0.23) {\tiny $\beta_2$};
        \node  at (0.65,0.2) {\tiny $\beta_4$};
        \node  at (0.75,0.2) {\tiny $\beta_5$};
    \end{scope}
\end{tikzpicture}
    \caption{Normalised magnetisation magnitudes $x_i=\frac{y_i}{a_i}$ as a function of $\be$. Here the relative sizes are $\{a_i\}=\{1,0.8,0.7,0.6,0.5\}$, with $k=5$. The are exactly four phase transitions: $\be_1\approx 2.2$, $\be_2\approx 3.1$, $\be_{4}= 10/3$, and $\be_{5}= 4$.}
    \label{fig:thm}
\end{figure}
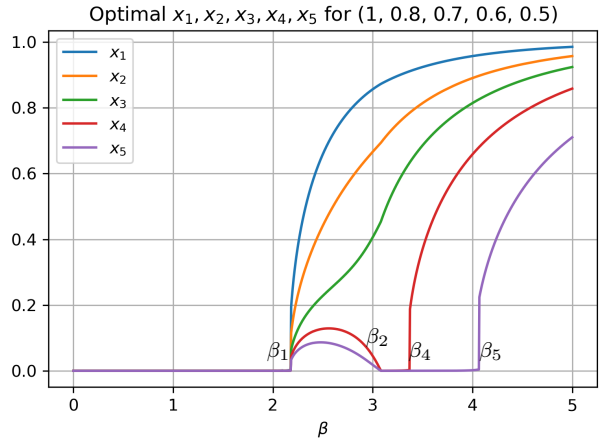

For the antiferromagnetic Heisenberg interaction for spin 1/2, with Hamiltonian given by equation (\ref{eq:H1}), we describe the set of $k$-exchangeable Gibbs states as a function of $\be$ and identify all phase transitions. See an example in Figure \ref{fig:thm}. We will focus on the case were there is a unique largest needle $a_1>a_2\geq \dots \geq a_k $. The case with two (or more) largest needles $a_1=a_2$ is easier to analyse and is treated in Remark \ref{remark:equalsize} at the end of Section \ref{Npartite}.

\begin{theorem}\label{thm:phase}
Consider the model given by equation (\ref{eq:H1}) with $k\geq 2$ and relative sizes $a_1> a_2\geq \dots \geq a_k$. Let $\be_\mathrm{1}\in (\frac{2}{a_1},\frac{2}{a_{2}})$ be the solution to: 
            \begin{equation}\label{eq:kpol}
    \sum_{i=1}^k \frac{1}{2-a_i\be_1}=\frac{k-1}{2}
\end{equation}
Furthermore if $a_1<\sum_{i=2}^k a_i $, define $\be_2\in (\frac{2}{a_3},\infty)$ by $\Delta y(\be_2)=0$. Otherwise set $\be_2=\infty$. 

The magnetisation magnitudes 
$y_i=y_i(\be)$ as functions of $\beta$ satisfy:
\begin{enumerate}[i]
    \item $y_i(\be)=0$ for $\be\leq \be_1$.
    \item For all $\be \in (\be_1,\be_2) $, $\Delta y >0$ and
    \begin{align}\label{eq:diff0}
    \begin{cases}
        \artanh \frac{y_1}{a_1} &=\frac{\be}{2}(y_1-\Delta y)\\
     \artanh \frac{y_i}{a_i} &=\frac{\be}{2} (y_i+\Delta y) ,\quad i\neq 1.
    \end{cases}
\end{align}
    
    In particular, there exist constants 
    $c_i>0$ such that:
    \begin{equation}\label{eq:critexp}
y_i(\be)=c_i\sqrt{\be-\be_1}+\mathcal{O}(\be-\be_1)^{\frac{3}{2}}
    \end{equation}
    \item For all $\be\in (\be_2,\infty)$, $\Delta y <0$ and 
    \begin{equation}\label{eq:low}
        \artanh \frac{y_i}{a_i} =\frac{\be}{2}y_i,\quad \forall i.
    \end{equation}
    In particular if $\be_2<\frac{2}{a_i}$, define $\be_i=\frac{2}{a_i}$ and we then have:
\begin{equation}\label{eq:crit2}
        y_i(\be)=a_i\sqrt{\frac{3a_i}{2}}\sqrt{\be-\be_i}+\mathcal{O}(\be-\be_i)^{\frac{3}{2}},\quad \be\geq \be_i
    \end{equation}
\end{enumerate}
\end{theorem}

Notice in Theorem \ref{thm:phase}, that the possible directions that the vectors $\vec{y}_i$ can have are determined by the three phases:  $y_i=0, \Delta y>0$ and $\Delta y<0$
separated by $\be_1$ and $\be_2$ respectively. In the phase $\Delta y >0$ all $\vec{y}_i$ lie on the same line with $\vec{y}_1$ being antiparallel to the rest. Here we have a net magnetisation equal to $\Delta y$. The vectors $\vec{{y}}_i$ are unique up to a rotation of $\vec{y}_1$. Thus the $k$-exchangeable extremal Gibbs state $\mathcal{G}_k$ can be represented by the unit sphere $S^2$. While in the phase $\Delta y<0$ the vectors $\vec{y}_i$ form a polygon $\sum_i \vec{y}_i$. Here there is no net magnetisation as they cancel each other. In this case $\mathcal{G}_k$ is represented by a subgroup of $S^1\times (S^2)^{\times (k-2)}$. 

\begin{figure}[h]
    \centering

\begin{minipage}{0.3\textwidth}
        \centering
        \begin{tikzpicture}[scale=0.8]
    \filldraw (0,0) circle (0.13cm) node[anchor=east]{1};
    \filldraw (4,0) circle (0.13cm) node[anchor=west]{2};
    \filldraw (0,4) circle (0.13cm)node[anchor=east]{3};
    \filldraw (4,4) circle (0.13cm) node[anchor=west]{4};
    \draw[thick, blue] (0,0)-- (2,0) node[anchor=north]{\Large $-$}--(4,0);
    \draw[thick, red] (4,0)-- (4,2) node[anchor=west]{\Large $+$}--(4,4);
    \draw[thick, blue] (0,0)-- (0,2) node[anchor=east]{\Large $-$}--(0,4);
    \draw[thick, red] (4,4)-- (2,4) node[anchor=south]{\Large $+$}--(0,4);
    \draw[thick, blue] (4,0)-- (2,2) node[anchor=west]{\Large $-$}--(0,4);
\end{tikzpicture}
    \end{minipage}
    \hfill
    \begin{minipage}{0.3\textwidth}
        \centering
        \begin{tikzpicture}[scale=0.8]
    \filldraw (0,0) circle (0.13cm) node[anchor=east]{1};
    \filldraw (4,0) circle (0.13cm) node[anchor=west]{2};
    \filldraw (0,4) circle (0.13cm)node[anchor=east]{3};
    \filldraw (4,4) circle (0.13cm) node[anchor=west]{4};
    \draw[thick, blue] (0,0)-- (2,0) node[anchor=north]{\Large $-$}--(4,0);
    \draw[thick, red] (4,0)-- (4,2) node[anchor=west]{\Large $+$}--(4,4);
    \draw[thick, blue] (0,0)-- (0,2) node[anchor=east]{\Large $-$}--(0,4);
    \draw[thick, red] (4,4)-- (2,4) node[anchor=south]{\Large $+$}--(0,4);
    \draw[thick, red] (4,0)-- (2,2) node[anchor=west]{\Large $+$}--(0,4);
\end{tikzpicture}
    \end{minipage}
    \hfill
    \begin{minipage}{0.3\textwidth}
        \centering
        \begin{tikzpicture}[scale=0.8]
    \filldraw (0,0) circle (0.13cm) node[anchor=east]{1};
    \filldraw (4,0) circle (0.13cm) node[anchor=west]{2};
    \filldraw (0,4) circle (0.13cm)node[anchor=east]{3};
    \filldraw (4,4) circle (0.13cm) node[anchor=west]{4};
    \draw[thick, blue] (0,0)-- (2,0) node[anchor=north]{\Large $-$}--(4,0);
    \draw[thick, red] (4,0)-- (4,2) node[anchor=west]{\Large $+$}--(4,4);
    \draw[thick, blue] (0,0)-- (0,2) node[anchor=east]{\Large $-$}--(0,4);
    \draw[thick, blue] (4,4)-- (2,4) node[anchor=south]{\Large $-$}--(0,4);
\end{tikzpicture}
    \end{minipage}

    \caption{Three examples of weights for $k=4$. If the weights correspond to $u_{ij} \neq 0$ (with $v_{ij}=0$), then only the left-most is frustration free. If instead the weights correspond to  $v_{ij} \neq 0$ (with $u_{ij}=0$), then only the right-most has frustration. }
    \label{fig:two}
\end{figure}
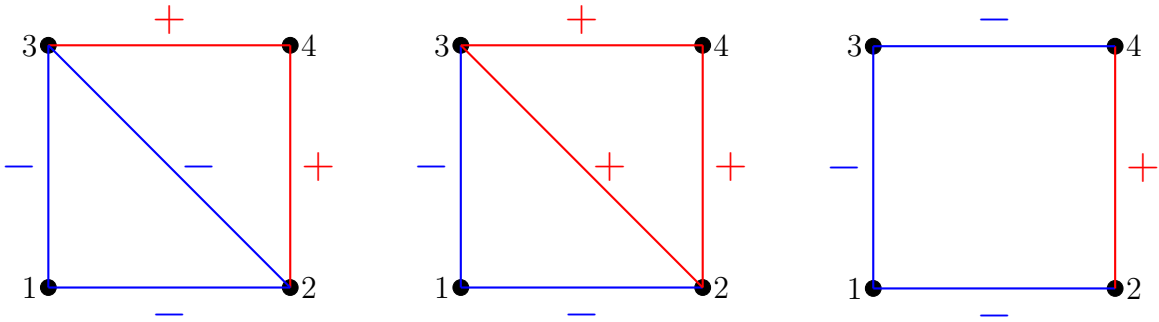

Another observation in Theorem \ref{thm:phase}, is that $\be_2$ is only implicitly defined. To find $\be_2$ in practice one needs to solve for $y_i(\be)$ in the transcendental equation (\ref{eq:low}) and then find the point where $\Delta y=0$. However $\be_1$ is given as a solution to a rational equation (\ref{eq:kpol}). Which can also be written as a $k$-degree polynomial. If we fix $a_1$, then increasing $k$ by partitioning the needles further causes $\be_1$ to approach closer to $\frac{2}{a_1}$.\newline\newline
For the O($2s+1$) invariant model, with Hamiltonian given by equation (\ref{eq:H2}), we will describe which parameters $u_{ij}$ and $v_{ij}$ lead to frustration. For this model we allow any spin number $s=\frac{1}{2},1,\frac{3}{2},\dots$. We define a graph $G=(V,E)$ where each vertex $i\in \{1,2,\dots,k\}=V$ represents a needle. Let $E$ be the pairs of needles $(i,j)$ with non-zero interaction $(u_{ij},v_{ij})\neq (0,0)$. We assume that $G$ is connected, meaning that all needles depend on each other in some way. The condition for frustration only depends on the sign $\pm$ of $u_{ij}$ and $v_{ij}$ respectively. So for each edge let  $(\operatorname{sgn} u_{ij},\operatorname{sgn} v_{ij})$ be its weight. Here we interpret $\operatorname{sgn}(0)$ as $\pm$.
\begin{theorem}\label{thm:graph}
  The model given by (\ref{eq:H2}) is frustration free, for all $\be$, if and only if there exists a partition of the connected graph $G=(V,E)$ into two subgraphs $(V_1,E_1)$ and $(V_2,E_2)$ and a bipartite graph $(V_1,V_2,E_{12})$:
  $$ V=V_1\sqcup V_2,\quad E=E_1\sqcup E_2 \sqcup E_{12}, $$
  such that either all weights in $E_1\sqcup E_2$ are $(--)$ and $E_{12}$ are $(++)$, or alternatively all weights in $E_1\sqcup E_2$ are $(-+)$ and $E_{12}$ are $(+-)$. 
\end{theorem}
Let us look at some small examples of Theorem \ref{thm:graph}. To simplify we look at models with only $T$-interaction ($v_{ij}=0$) or only $Q$-interaction $(u_{ij}=0)$. First, consider $k=3$. Here all models with $u_{ij}=0$ are frustration free. While if all $v_{ij}=0$ the signs of $u_{ij}$ must be $---$ or $-++$. Figure \ref{fig:two} shows three examples for $k=4$.


\section{Antiferromagnetic Heisenberg model}\label{Npartite}
In this Section we will prove Theorem \ref{thm:phase}. We assume that the relative sizes are ordered $a_1>a_2\geq \dots \geq a_k$ with $a_1$ being the unique largest. We start by reducing the problem to a minimisation problem in real analysis. The variables we minimise over will be the magnetisation vectors $\vec{y}_i\in \mathbb{R}^3$ with $y_i\leq a_i$ defined in (\ref{eq:tau}). Also recall that $y_i=a_ix_i$. The extremal $k$-exchangeable Gibbs state $\tau=(\tau_1,\dots,\tau_k)\in \mathcal{G}_k$ is the argmin of the specific free energy (\ref{eq:free}). We use the parametrisation of $\tau_i$ given by (\ref{eq:tau}), and we complete the square in the energy term. Finding the $\mathcal{G}_k$ is now equivalent to finding the minimiser $\vec{y}_i\in \mathbb{R}^3$ such that $ y_i:=|\vec{y}_i|\leq a_i$ of:
\begin{equation}\label{eq:free_square}
    f_\be(\vec{y})= \frac{1}{4}\big|\sum_{i=1}^k \vec{y}_i\big|^2 +\frac{1}{\beta}\sum_{i=1}^k  a_ig_{\be a_i}(x_i)
\end{equation}
where
\begin{equation}\label{eq:g}
     g_{\theta}(x)= \ln{(1+x)}-(1-x)\artanh x-2\ln 2 -\frac{\theta x^2}{4} 
\end{equation}
    
Notice that
\begin{equation}\label{eq:tay0}
    g_\theta(x)=-2\ln 2 +x^2(\frac{1}{2}-\frac{\theta}{4})+\frac{x^4}{12}+\mathcal{O}(x^6),\quad \forall x\in[0,1]
\end{equation}
The term $|\sum_i \vec{y_i}|^2$ means that the vectors $\vec{y}_i$ try to form a polygon for every choice of magnitudes $y_i=a_ix_i$.

\begin{lemma}\label{lemma:minfunc} 
The function $g_{\theta}(x)$ has a unique local minimum given by:
    \begin{equation}
       x_\theta^*:=\argmin_{x\in [0,1]} g_\theta(x)=\begin{cases} 0,\quad \theta  \leq 2.\\[0.3em]
        x \, : \, \artanh{x}=\frac{\theta x}{2}>0,\quad \theta  > 2. 
        \end{cases}
    \end{equation}
    Moreover $x_\theta^*$ is increasing in $\theta$ and the critical exponent at $\theta=2$ is $\frac{1}{2}$:
\begin{equation}\label{eq:crit}
x_{\theta}^*=\sqrt{\frac{3}{2}}\sqrt{\theta-2}+\mathcal{O}(\ta-2)^{\frac{3}{2}},\quad \ta\geq 2
\end{equation}
\end{lemma}
\begin{Proof} Differentiate:
\begin{equation}\label{eq:tay}
     g_{\theta}'(x)=\artanh(x)-\frac{\theta x}{2}=x\Big(1-\frac{\theta}{2}\Big)+\frac{x^3}{3}+\frac{x^5}{5}+\mathcal{O}(x^7)
\end{equation}
For $\theta  \leq 2$, the derivative is positive for all $x>0$ thus $x_\theta^*=0$. While for $\theta>2$ let $x_{\theta}^*>0$ be the unique positive zero to $g'_\theta(x)=0$. This is indeed the unique minimum since the derivative is negative for $(0,x_\theta^*)$. Notice also that $x_\theta^*$ is strictly increasing for $\theta\geq 2$ since 
$ \frac{\artanh x}{x} $ is strictly increasing in $x$. We obtain (\ref{eq:crit}) by keeping the qubic term in the Taylor expansion (\ref{eq:tay}) and solve for $x$:
$$ 0 = x\Big(1-\frac{\theta}{2}\Big)+\frac{x^3}{3} $$
\end{Proof}
Also let us denote $y_{\be a_i}^*=a_ix_{\be a_i}^*$. If $y_i=y^*_{\be a_i}$ we can think of needle $i$ as happy. And unhappy if $y_i\neq y^*_{\be a_i}$.

\begin{Proof}[Proof of Theorem 2.]
For $\be\leq \frac{2}{a_1}$ the free energy is minimised term-wise by $x_i=0$, see Lemma \ref{lemma:minfunc}. For the rest of this proof assume that $\be>\frac{2}{a_1}$. \newline\newline
We will consider the three different cases of $\Delta y= y_1- y_2-\dots-y_k$ being zero, positive and negative. We will see that these are the three phases we are looking for.\newline\newline
First we prove that $\Delta y <0$ implies $y_i=y_{\be a_i}^*$ for all $i$. 

Notice that $y_1\geq y_i$ because the unique minimisers of $g_{\be a_i}$ satisfy $y_{\be a_1}^*> y_{\be a_i}^*$, see Lemma \ref{lemma:minfunc}. Then $\Delta y <0$ implies that the vectors $\vec{y}_i$ form a polygon. And the free energy takes the form:
$$ f_{\be}(y)=\frac{1}{\be}\sum_{i=1}^k a_i g_{\be a_i}(x_i) $$
By definition, $y_i=y_{\be a_1}^*$ for all $i$.

Secondly we show that the phase $\Delta y<0$ occurs if and only if $a_1<a_2+\dots +a_k$. Furthermore if true, $\Delta y<0$ occurs in an interval $\be\in (\be_2,\infty)$ for some $\be_2\in (\frac{2}{a_3},\infty)$. 

If $a_1\geq a_2+\dots +a_k $, the fact $x_{\be a_1}^*\geq x^*_{\be a_i} $ implies that $\Delta y \geq 0 $. If $a_1<a_2+\dots +a_k$, let $\be$ be sufficiently large. Then $\Delta y <0$ because $x_{\be a_i}^*\rightarrow 1$ as $\be\rightarrow\infty$. 

By Lemma \ref{lemma:minfunc}, we know that $\De y<0$ is equivalent to: 
$$ \artanh x_1 < \artanh{x_2}+\dots +\artanh{x_k} $$
And $x_2=x_{\be a_2}^*$ increase faster then $x_1=x_{\be a_2}^*$ because $x_{\theta}^*$ is concave for $\theta\in [2,\infty)$. This shows that the phase $\Delta y<0$ occurs in an interval. And $\be_2$ is the unique solution to:
\begin{equation}\label{be2}
    \artanh x_{\be a_1}^* = \artanh{x_{\be a_2}^*}+\dots +\artanh{x_{\be a_k}^*}
\end{equation}
The fact that $\be_2>\frac{2}{a_3}$ follows from $y_{\be a_1}^*>y_{\be a_2}^*$ for $\be \in (\frac{2}{a_2},\infty)$. Equation (\ref{eq:crit2}) follows directly from equation (\ref{eq:crit}) in Lemma \ref{lemma:minfunc}.
\newline\newline
Now we focus on the positive phase $\Delta y>0$. We prove that $\Delta y >0$ occurs in an interval $\be\in (\be_1,\be_2) $, where $\be_1$ is given by (\ref{eq:kpol}).

With $\Delta y>0$ the free energy takes the form:
$$  f_\be(y)= \frac{1}{4}(\Delta y)^2 +\frac{1}{\beta}\sum_{i=1}^k  a_ig_{\be a_i}(x_i) $$
Differentiate:
\begin{align}\label{eq:diff}
    \begin{cases}
    \frac{\partial f}{\partial y_1}&=\frac{1}{\be}\artanh x_1 -\frac{1}{2}(y_1-\Delta y)\\ 
     \frac{\partial f}{\partial y_i}&=\frac{1}{\be}\artanh x_i -\frac{1}{2} (y_i+\Delta y) ,\quad i\neq 1.
     \end{cases}
\end{align}
From (\ref{eq:diff}) we see that $x_i>0$. If not we would have the contradiction $\frac{\partial f}{\partial y_i}<0$ and $y_i=0$. With $x_i>0$ the derivatives must be zero.
\begin{align}\label{eq:diff3}
\begin{cases}
    \artanh x_1 &=\frac{\be}{2}(y_1-\Delta y)\\
     \artanh x_i &=\frac{\be}{2} (y_i+\Delta y) ,\quad i\neq 1.
     \end{cases}
\end{align}
The solution of this system is analytic with respect to $\be$. To see this, linearise the equation and solve the linear system of equations with Gaussian elimination. The only issue that can occur is division by zero but we know the solution is bounded. 

Let us prove that the phase ends with $\be_2$. Notice that $x_1$ is increasing and thus non-zero in this phase. If there ever is a $\be$ such that $\Delta y = 0 $, then from (\ref{eq:diff3}) we see that $y_i=y_{\be a_i }^*$. From equation (\ref{be2}) we realise that this $\be$ is precisely $\be_2$.

Next let us find the beginning of the phase $\be_1$. If $\Delta y \rightarrow 0$ with $y_1\not\rightarrow 0$ we arrive at $\be_2$. So try with $y_1\rightarrow 0$. We also have $y_i\rightarrow 0$ since $\Delta y >0$. We can now linearise system (\ref{eq:diff3}):
\begin{equation}\label{eq:system}
 \begin{cases}
     y_1=\frac{\be a_1}{2} (y_1-\Delta y) \vspace{0.3cm} \\ 
      y_i=\frac{\be a_i}{2}(y_i+\Delta y), \quad i\neq 1.
 \end{cases}
 \end{equation}  
If $\be$ satisfies (\ref{eq:system}) it is easy to check that (\ref{eq:eq}) holds: 
\begin{equation}\label{eq:eq}
    \sum_{i=1}^k \frac{1}{2-a_i\be}=\frac{k-1}{2}
\end{equation} 
This equation has a unique solution in $\be\in (\frac{2}{a_1},\frac{2}{a_2})$. But we still need to show that $\be_1<\frac{2}{a_2}$. We do this by showing that $\Delta y >0$ for $\be\in (\frac{2}{a_2}-\ep,\frac{2}{a_2})$ with $\ep$ small enough. 

Firstly show that $x_1>0$ for $\be \in (\frac{2}{a_2}-\ep,\frac{2}{a_2})$. Compare all $y_i=0$ with $y_1=\ep$ and $\vec{y}_2=-\vec{y}_1$. Since the vectors cancel each other, only $g_{\be a_1}(\ep)$ and $g_{\be a_2}(\ep)$ needs to be analysed. Use the Taylor expansion of given by (\ref{eq:tay0}).
The difference in free energy is then:
$$ \frac{a_1}{4\be}\ep^2(2-a_1\be)+\mathcal{O}(\ep^3),\quad \forall \be\in  (\frac{2}{a_2}-\ep,\frac{2}{a_2}) $$
Which is negative for small $\ep$. Notice that the contribution from $g_{\be a_2}(\ep)$ is absorbed in $\mathcal{O}(\ep^3)$.

Secondly show that if $y_1>0$ then $\Delta y>0$ for $\be\in (\frac{2}{a_1},\frac{2}{a_2}) $. To see this start with $\Delta y=0$ and $y_1,y_i>0$. Decrease $y_i$ with $\ep$ to $y_i'=y_i-\ep$. The change in free energy is:
$$ f_\be(y_i-\ep)-f_\be(y_i) = \mathcal{O}(\ep^2)-\ep \Big(x_i(1-\frac{\be a_i}{2})+\frac{x_i^3}{3}\Big) $$
Which is negative for small enough $\ep$.

To find equation (\ref{eq:critexp}), let us keep the cubic term in the Taylor expansion in (\ref{eq:diff3}).
\begin{equation}
    \begin{cases}
        \frac{2}{3}x^3_1=(\be -\be_1)(y_1-\Delta y)\\ 
         \frac{2}{3}x^3_i=(\be -\be_1)(y_i+\Delta y)
    \end{cases}
\end{equation}
We can expand $y_i$ in powers of $(\be-\be_1)$. By consistency we must have (\ref{eq:critexp}). \newline\newline
Finally consider the last phase $\Delta y =0$. We know that $\Delta y >0$ in $(\be_1,\be_2)$ and $\Delta y<0$ in $(\be_2,\infty)$. Thus we must have $\Delta y =0$ in $(\frac{2}{a_1},\be_1]$. From the latest argument we know that $\Delta y =0$ implies $y_1=0$. But then all $y_i=0$. This proves Theorem \ref{thm:phase}.
\end{Proof}

\begin{remark}\label{remark:equalsize}
   Let us consider what happens in the less interesting case $a_1=a_2\geq a_3 \geq \dots \geq a_k$. Let all $x_i$ minimise $g_{\be a_i}(x_i)$, as given by Lemma \ref{lemma:minfunc}. Then by definition the second term in the free energy (\ref{eq:free_square}) is minimised. It is then sufficient to check that the first term $\frac{1}{4}|\sum_i \vec{y}_i|^2$ will be zero. Notice that $y_1=y_2\geq y_3\geq \dots \geq y_k $. Then it is easy to choose the directions to set $\sum_i \vec{y}_i=0$. 
\end{remark}

\section{Orthogonally invariant models}
In this Section we prove Theorem \ref{thm:graph}.  Just like in the previous Section,
consider extremal $k$-exchangeable states $\rho$ given by (\ref{eq:state}) and Theorem \ref{thm:definetti}. However now we work with a general spin number $s$. Notice that the interaction terms can simplified as follows:
$$ \tr (\tau_1\otimes \tau_2 T_{12})=\tr \Big(\tau_1\otimes \tau_2 \sum_{\al,\be=-s}^s |\al\be\rangle\langle \be\al|\Big) = \tr \tau_1 \tau_2 $$
$$ \tr (\tau_1\otimes \tau_2 Q_{12})=\tr \Big(\tau_1\otimes \tau_2 \sum_{\al,\be=-s}^s |\al\al\rangle\langle \be\be|\Big) = \tr \tau_1 \tau_2^\mathrm{T} $$
The Gibbs state of model (\ref{eq:H2}) is the
argmin of the specific free energy (\ref{eq:free}):
\begin{equation}\label{eq:freeO}
    f_\be(\tau)= \sum_{i\neq j }^ka_ia_j\tr\Big(u_{ij}\tau_i\tau_j  +v_{ij}\tau_i\tau_j\T \Big)+\frac{1}{\beta}\sum_{i=1}^k a_i\tr(\tau_i\ln \tau_i)
\end{equation}
We diagonalise $\tau_i$ as follows $\tau_i=U_i^\dagger \operatorname{diag}(\la_i)U_i$. Here $U_i$ is unitary and assume that the vector of eigenvalues $\la_i=(\la_{i,-s},\la_{i,1-s},\dots,\la_{i,s-1},\la_{i,s})$ is written in increasing order. Let us introduce two matrices $P$ and $Z$, where $P$ denote the anti-diagonal matrix with ones on the anti-diagonal and where $Z$ is a symmetric unitary matrix such that $Z^2=P$. For instance for $2s+1=5$ we can take:
$$ P=\begin{pmatrix}
    & & & & 1\\
    & & & 1 &\\
    & & 1 & &\\
    & 1 & & &\\
    1 & & & &\\
\end{pmatrix},\quad  Z= \frac{1}{\sqrt{2i}}\begin{pmatrix}
    1 &  &  &  & i \\
     & 1 &  & i &  \\
     &  & \sqrt{2i} &  &  \\
     & i &  & 1 &  \\
    i &  &  &  & 1 \\
\end{pmatrix} $$
The difficult part of proof is the "only if" direction, so we will need three lemmas to prepare for that. 

\begin{lemma}\label{lemma:bounds}
    A model (\ref{eq:H2}) is  frustration free if and only if we have for all $i,j$:
    \begin{equation}\label{eq:condlemma}
        \tr(\tau_i\tau_j)=\begin{cases} \la_i\cdot \la_j, \quad \text{if } u_{ij}<0\\
        \la_i \cdot P\la_j,\quad \text{if } u_{ij}>0
    \end{cases} \quad \text{and}\quad 
    \tr(\tau_i\tau_j\T)=\begin{cases} \la_i\cdot \la_j, \quad \text{if } v_{ij}<0\\
        \la_i \cdot P\la_j,\quad \text{if } v_{ij}>0 
        \end{cases}
    \end{equation}
\end{lemma}
\begin{Proof}
    For a frustration free model we know that all $(u_{ij}\tr \tau_i\tau_j+v_{ij}\tr{\tau_i\tau_j}\T)$ are minimised with fixed $\la_i$. We can rewrite $\tr(\tau_i\tau_j)=\la_i \cdot |U_iU_j|^2 \la_j $, where $|U_iU_j|^2$ is a doubly stochastic matrix. By Birkoff's theorem and the rearrangement inequality we have the bounds:
\begin{equation}\label{eq:bounds}
\begin{cases}
    \la_i \cdot P\la_j\leq \tr \tau_i \tau_j\leq \la_i \cdot \la_j\\
    \la_i \cdot P\la_j\leq \tr \tau_i \tau_j\T \leq \la_i \cdot \la_j
\end{cases}   
\end{equation}
We claim that any pair of equalities can be attained simultaneously. The upper equalities can be attained by $U_i,U_j=I$. The lower equalities can be attained by $U_i=P$ and $U_j=I$. To get $\tr \tau_i\tau_j=\la_i\cdot \la_j$ and $\tr \tau_i\tau_j\T=\la_i\cdot P\la_j$, take $U_i,U_j=Z$. And to get $\tr \tau_i\tau_j=\la_i\cdot P\la_j$ and $\tr \tau_i\tau_j\T=\la_i\cdot \la_j$, take $U_i=Z$ and $U_j=\overline{Z}$. 
\end{Proof}
With help of Lemma (\ref{lemma:bounds}), it is now possible to write down the free energy only in terms of eigenvalues. This allows us to show in the following lemma that the eigenvalues must be on a very specific form.

\begin{lemma}\label{lemma:contra}
    If a model (\ref{eq:H2}) is  frustration free and cannot be partitioned into $V_1$ and $V_2$ as in Theorem \ref{thm:graph}, then, for large enough $\be$, there exists a positive integer $l\leq \lceil s\rceil $, such that all $\la_i$ have precisely $l$ largest values and $l$ smallest values.
\end{lemma}

\begin{Proof}
By Lemma \ref{lemma:bounds}, the free energy $f_\be$ in equation (\ref{eq:freeO}) can be written in terms of $\la_i\cdot P\la_j$ and $\la_i\cdot \la_j$. Split $u_{ij}$ and $v_{ij}$ into positive and negative part: $u_{ij}=u_{ij}^++u_{ij}^-$ and $v_{ij}=v_{ij}^++v_{ij}^-$. Also denote $w^{\pm}_{ij}=a_ia_j(u_{ij}^{\pm} +v_{ij}^{\pm})$. Then for frustration free models:
\begin{equation}
     f_\be(\la)= \sum_{i\neq j }^k\la_i\cdot\big(w_{ij}^+ P  +w_{ij}^-  \big)\la_j+\frac{1}{\beta}\sum_{i=1}^k a_i\la_i\cdot\ln \la_i
\end{equation} 
Next let us show two properties of how repeated eigenvalues $\la_i=(\la_{i,-s},\dots,\la_{i,m},\dots,\la_{i,s})$ transfers to other vectors $\la_j$.

\textbf{Claim 1.} 
\textit{If $w_{ij}^-<0$ we have: 
$$ \la_{i,m}=\la_{i,m+1} \quad \Leftrightarrow \quad \la_{j,m}=\la_{j,m+1}$$
and if $w_{ij}^+>0$ we have:}
$$ \la_{i,m}=\la_{i,m+1} \quad \Leftrightarrow \quad \la_{j,-m}=\la_{j,-m-1} $$

\begin{Proof}
Consider $w_{ij}^+>0$ and $m=s$, the other cases are done similarly. Suppose the opposite that $\la_i=(\dots,y,z)$ where $y<z$ and $\la_j=(x,x,\dots)$. Firstly, $x>0$ for all $\be<\infty$, since the derivative of the entropy is infinite at $x=0$. Compare the free energy of $\la_j$ with $(x-\ep,x+\ep,\dots)$ for some small $\ep>0$. Then the energy in the interaction $w_{ij}^+\la_i \cdot P \la_j$ is changed by $\ep w_{ij}^+(y-z)<0$. The energy in all other interactions is either unchanged or lowered since $(w_{ij}^+P+w_{ij}^-)\la_j$ is decreasing. The entropy is decreased by $\frac{a_j}{\be}\frac{1}{x}\ep^2+\mathcal{O}(\ep^4)$. The total change in free energy is:
$$ f_\be(x-\ep,x+\ep)-f_\be(x,x)\leq \ep w_{ij}^+(y-z)+\frac{a_j}{\be}\frac{1}{x}\ep^2+\mathcal{O}(\ep^4) $$
Which is negative for small enough $\ep$, which is a contradiction.
\end{Proof}
\textbf{Claim 2.}\textit{
    If $\la_i$ has precisely $l$ smallest and largest values, for some $i$, then that property is true for all $i=1,2,\dots,k$.}
    \begin{Proof}
    If $\la_i$ has precisely $l$ smallest values, then, by Claim 1, all neighbours $j$ with $w_{ij}^-<0$ also have $l$ smallest values while all neighbours with $w_{ij}^+>0$ have $l$ largest values. If $\la_i$ has both $l$ smallest and $l$ largest, this property transfers to all neighbours. And since the graph is connected this property must transfer to all $1,2,3,\dots,k$.
\end{Proof}
Consider the set $S$ of needles $i$ such that $w_{ij}^-<0$ for some $j$. In the limit $\be\rightarrow\infty$ the free energy approaches $\sum_{i\neq j} w_{ij}^-<0$, which can only happen if $\la_i$ has a unique largest value for all $i\in S$.

Try to colour every needle in $V$ black or white such that: all needles in $S$ are white and if $w_{ij}^+>0$ then $i$ and $j$ have different colours. 

If such a colouring exists then let $V_1$ be the white needles and $V_2$ be the black needles. Then the weights in $E_1,E_2$ is $(--)$ and the weights in $E_{12}$ is $(++)$. This is a partition allowed by Theorem \ref{thm:graph}, and thus a contradiction. 

Hence such a colouring does not exists. If $S$ is non-empty, then there exists a path with $w_{ij}^+>0$ of odd steps from $S$ to $S$. Using Claim 1 on this path implies that some $i\in S$ have a unique smallest value as well. By Claim 2, all needles  $i\in V$ have a unique largest and smallest value. So here $l=1$.\newline \newline
If instead $S$ is empty and there is no colouring, there must exists an odd cycle with $w_{ij}^+>0$. By Claim 1, any $\la_i$ on this cycle must be on the form:
\begin{equation}\label{eq:form}
    \la_i=(\underbrace{b_i,\dots,b_i}_{l},c_i,\dots,d_i,\underbrace{e_i,\dots,e_i}_{l})
\end{equation}
with $b_i<c_i\leq d_i<e_i$ and $1\leq l \leq \lceil s \rceil $ or all $\la_i$ on this cycle is on the form $\la_i=(\frac{1}{2s+1},\dots,\frac{1}{2s+1})$. We can exclude the second case for large enough $\be$, since then free energy would approach $\frac{1}{2s+1}\sum_{i\neq j}w_{ij}^+$ as $\be\rightarrow\infty$, but we know the correct answer is $f_\be\rightarrow 0$. By Claim 2, all $i=1,2,\dots,k$ are on the form (\ref{eq:form}) for large $\be$.
\end{Proof}

\noindent With the knowledge from Lemma \ref{lemma:contra}, it is reasonable to define an equivalence relation for unitary matrices:
   \begin{equation}
       A\sim_l B\quad \Leftrightarrow\quad |AB^\dagger|^2=\begin{pmatrix}
           X_l & 0 & 0\\
           0 & Y_{2s+1-2l} & 0\\
           0 & 0 & Z_l
       \end{pmatrix}
   \end{equation}
   Here $X_l,Y_{2s+1-2l},Z_l$ are arbitrary doubly stochastic matrices of size $l,2(s-l)+1$ and $l$, where $1\leq l\leq \lceil s\rceil $. And $|\cdot|^2$ denotes element-wise absolute value square. Notice that the element-wise operation does not change the block structure. It is easy to see that $\sim_l$ is well-defined and is compatible with multiplication by unitary matrices and complex conjugation:$$\overline{A}\sim_l \overline{B}\quad \Leftrightarrow \quad A\sim_l B\quad \Leftrightarrow\quad AC\sim_l BC$$

\begin{lemma}\label{lemma:cond}
    If a model (\ref{eq:H2}) is frustration free and cannot be partitioned into $V_1$
and $V_2$ as in Theorem \ref{thm:graph}, then, for large enough $\be$, there exists an integer $l$ and unitary matrices $U_i$ in (\ref{eq:freeO}) such that for all $i,j$:
    \begin{equation}\label{eq:cond}
    \begin{cases}
        U_i\sim_l   U_j,\quad \text{if }u_{ij}<0,\\
        U_i\sim_l PU_j \quad \text{if }u_{ij}>0.
    \end{cases} \quad \text{and}\quad \begin{cases}
        U_i\sim_l \overline{U}_j,\quad \text{if }v_{ij}<0,\\
        U_i\sim_l P\overline{U}_j \quad \text{if }v_{ij}>0.
    \end{cases}
\end{equation}
\end{lemma}
\begin{Proof}
Assume $\be$ is large enough so we can define $l$ as in Lemma \ref{lemma:contra}. We assume a frustration free model, so we have conditions (\ref{eq:condlemma}) in Lemma \ref{lemma:bounds}. 

If $u_{ij}<0$ then $ \tr \tau_i\tau_j = \la_i\cdot\la_j $. Rewrite this equation:
$ \la_i |U_i U_j^\dagger|^2\la_j=\la_i\cdot\la_j $. By the rearrangement inequality, the $l$ largest and smallest values in $\la_j$ must stay in their positions after the mixture of permutations $|U_i U_j^\dagger|^2$ is applied to $\la_j$. This can only happen if $U_i\sim_l U_j$.

If instead $u_{ij}>0$ then $ \la_i |U_i U_j^\dagger|^2\la_j = \la_i\cdot P\la_j $. By the rearrangement inequality, the $l$ largest and smallest values in $\la_j$ must switch with each other after the mixture of permutations $|U_i U_j^\dagger|^2$ is applied to $\la_j$. This can only happen if $U_i\sim_l PU_j$.

Similar analysis for $v_{ij}$ leads to (\ref{eq:cond}). The only difference is that $|U_iU_j^\dagger|^2$ is replaced by $|U_iU_j\T|^2$.
\end{Proof}
Lemma (\ref{lemma:cond}) describe how all the unitary matrices $U_i$ must be related to each in order to give a frustration free model. It is the main tool we need to prove Theorem \ref{thm:graph}.

\begin{Proof}[Proof of Theorem \ref{thm:graph}] 
We start by showing that frustration free and not having the partition $V_1\sqcup V_2$ as in Theorem \ref{thm:graph} is impossible. Assume conditions in Lemma \ref{lemma:cond} so (\ref{eq:cond}) hold. 

Let us denote $A=U_1$ for clarity. Since our graph $G=(V,E)$ is connected there is a path from 1 to any $i\in V$. Using condition (\ref{eq:cond}) on this path results in some sequence of multiplication with $P$ and complex conjugation. For instance say $u_{12}>0$ and $v_{23}>0$ then: $U_1\sim_l PU_2\sim_l P(P\overline{U_3})=\overline{U}_3$. And in general, since multiplication by $P$ and complex conjugations are involutions that commute, there are only four possible equivalence classes:
$$ U_i\sim_l \,A,\, PA,\, \overline{A} \,\text{ or }\, P\overline{A}, \quad \forall i\in V $$
Let us divide into three cases: $A\sim_l \overline{A}$ or $ A\sim_lP\overline{A}$ or neither.

If $A\sim_l \overline{A}$, then there are only two equivalence classes $A$ and $PA$. Notice that $A\sim_l \overline{A}$ implies $\overline{U}_i\sim_l U_i$. In other words, condition (\ref{eq:cond}) is the same for $u_{ij}$ and $v_{ij}$. If $U_i\sim_l U_j $ then $u_{ij},v_{ij}\leq 0$. And if $ U_i\not  \sim_l U_j $ then $u_{ij},v_{ij}\geq 0$. Our graph can be partitioned into $V_1$ and $V_2$ being the two equivalence classes. The corresponding weights are $(--)$ in $E_1\sqcup E_1$ and $(++)$ in $E_{12}$.

If instead $A\sim_l P\overline{A}$, then the only equivalence classes are $A$ and $\overline{A}$. Notice that $A\sim_l P\overline{A}$ implies $\overline{U}_i\sim_l PU_i$. In other words, condition (\ref{eq:cond}) is the opposite for $u_{ij}$ and $v_{ij}$. If $U_i\sim_l U_j$ then $u_{ij}\leq 0$ while $v_{ij}\geq 0$. And if $U_i\not\sim_l U_j$ then $u_{ij}\geq 0$ while $v_{ij}\leq 0$. Our graph can be partitioned into $V_1$ and $V_2$ being the two equivalence classes. The corresponding weights are $(-+)$ in $E_1\sqcup E_1$ and $(+-)$ in $E_{12}$. \vspace{0.2cm}

\noindent 
\begin{minipage}[h]{0.48\textwidth}
    \hspace{0.4cm}The third case is that neither relation is fulfilled: $A\not\sim_l P\overline{A}$ and $A\not\sim_l \overline{A}$. From condition (\ref{eq:cond}) we see that only one of $u_{ij}$ and $v_{ij}$ can be non-zero. Figure \ref{fig:four} shows how the four equivalence classes are related by weights. If $U_i\sim_l U_j$ then $u_{ij}\leq 0$. If $U_i\sim_l PU_j$ then $u_{ij}\geq 0 $. If $U_i\sim_l \overline{U}_j$ then $v_{ij}\leq 0$. If $U_i\sim_l P\overline{U}_j$ then $v_{ij}\geq 0$. Here our graph can for instance be partitioned so that $V_1$ correspond to the classes $A,\overline{A}$ and $V_2$ correspond to the classes $PA,P\overline{A}$. The corresponding weights are $(--)$ in $E_1\sqcup E_1$ and $(++)$ in $E_{12}$.
\end{minipage}
\hfill
\begin{minipage}[h]{0.46\textwidth}
    \centering
   \begin{tikzpicture}{scale=0.9}
    \filldraw[blue] (0,0) circle (0.13cm);
    \draw (0,0)  node[anchor=east]{$P\overline{A}$};
    \filldraw[blue] (4,0) circle (0.13cm);
    \draw (4,0)  node[anchor=west]{$PA$};
    \filldraw[blue] (0,4) circle (0.13cm);
    \draw (0,4) node[anchor=east]{$A$};
    \filldraw[blue] (4,4) circle (0.13cm);
    \draw (4,4) node[anchor=west]{$\overline{A}$};
     \draw[very thick, red] (0,0)--(4,4);
    \draw[very thick, red] (4,0)--(0,4);
    \draw[very thick, blue,dashed] (0,0)--(4,0);
    \draw[very thick, blue,dashed] (0,4)--(4,4);
    \draw[very thick, red,dashed] (0,0)--(0,4);
    \draw[very thick, red,dashed] (4,0)--(4,4);
    \draw (2,4) node[anchor=south]{$v\leq 0$};
    \draw (2,0) node[anchor=north]{$v\leq 0$};
    \draw (4,2) node[anchor=west]{$v\geq 0$};
    \draw (0,2) node[anchor=east]{$v\geq 0$};
    \draw (2,1.6) node[anchor=north]{$u\geq 0$};
\end{tikzpicture} 
    
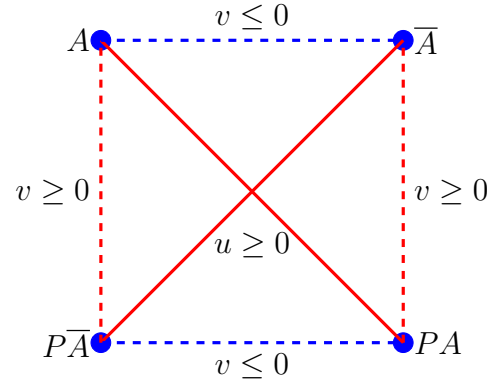
\captionof{figure}{Blue is negative and red is positive. Solid line and circle is $T$-interaction, and dashed line is $Q$-interaction.}
    \label{fig:four}
\end{minipage}\newline\newline
\noindent 
Finally we check that there is a frustration free construction if the graph $G=(V,E)$ can be partitioned as described in Theorem (\ref{thm:graph}). We need to see that (\ref{eq:condlemma}) is fulfilled. Let $i\in V_1$ and $j\in V_2$. If edges in $E_{12}$ have weight $(++)$ take $U_i=I$ and $U_j=P$. If instead edges in $E_{12}$ have weight $(+-)$ take $U_i=Z$ and $U_j=\overline{Z}$.

\end{Proof}

\appendix
\section{Quantum de Finetti theorem}\label{appendix}
In this Chapter we describe how to prove Theorem \ref{thm:definetti}. We follow the method presented by Caves, Fuchs and Schack for $k=1$ in \cite{Caves}. Focus on $k=2$ and $\mathcal{H}_1=\mathcal{H}_2=\mathbb{C}^2$. The idea is to map the matrix $\rho_n $ to a probability distribution and use combinatorics to decouple the probability distribution to a mixture of independent ones. Define four matrices  $E_\al$ by
$$ E_\al =\frac{1}{2}(1+\vec{v}_\al\cdot\vec{\si}),\quad \al=1,2,3,4. $$
where $\vec{v}_\al$ are the corners of a tetrahedron on the unit sphere. We will view $\al$ as an outcome $\{1,2,3,4\}$ from a probability experiment. Let $\vec{\al}=(\al_{1},\dots,\al_{n_1})$ and $\vec{\be}=(\be_{1},\dots,\be_{n_2})$ be vectors of such outcomes.  
Define a probability distribution $p_n=p_n(\vec{\al},\vec{\be})$ by
\begin{equation}\label{eq:n1}
    p_n(\vec{\al},\vec{\be}) :=\tr \Big(\rho_n\bigotimes_{i=1}^{n_1} E_{\al_i}\otimes \bigotimes_{j=1}^{n_2} E_{\be_j} \Big)
\end{equation}
Since $\rho_n$ is exchangeable we can extend it to $\rho_m$ as in the definition with $m_1\geq n_1$ and $m_2\geq n_2$. A nice combinatorial proof of the classical de Finetti is given by Heath and Sudderth in \cite{Heath}. We outline the idea here. From $\rho_m$ define a probability distribution $p_m(\vec{\ga},\vec{\de})$, where $\vec{\ga}=(\ga_{1},\dots,\ga_{m_1})$ and $\vec{\de}=(\de_{1},\dots,\de_{m_2})$ are vectors of outcomes. For $i=1,2,3,4$ let $c_i$ and $d_i$ be the number of times that $i$ occur in $\vec{\ga}$ and $\vec{\de}$ respectively. As a shorthand let $c=(c_1,c_2,c_3,c_4)$ and $d=(d_1,d_2,d_3,d_4)$. Use law of total probability and condition on $c$ and $d$ to conclude that
\begin{equation}
    p_n(\vec{\al},\vec{\be})=\sum_{c,d}p_m(\vec{\al},\vec{\be}|c,d)\mathbb{P}_m(c,d)
\end{equation}
where $\mathbb{P}_m(c,d)$ is the probability of getting frequencies $c,d$. Since $\rho_m$ is $k$-exchangeable, $p_m(\vec{\ga},\vec{\de}|c,d)$ will have a permutation symmetry $S_{m_1}\times S_{m_2}$. In particular, there is a uniform probability of all permutations of $\ga$ given $c$ and this fact is independent of $d$. In other words the probability factors into a product:
$$p_m(\vec{\ga},\vec{\de}|c,d) =p_m(\vec{\ga}|c) p_m(\vec{\de}|d) $$
To find $p_m(\vec{\ga}|c) $  uniquely label all $m_1$ symbols with indices. Since every symbol is unique the number of ways to get $c_{1}$ ones at the correct places in $\vec{al}$ will be the falling factorial $(c_{1})_{a_{1}}=c_{1}(c_{1}-1)\dots(c_{1}-a_{1}+1)$. In total we get:
$$ p_n(\vec{\al},\vec{\be})=\sum_{c,d}
 \frac{\prod_{i=1}^4(c_{i})_{a_{i}}(d_i)_{b_i}}{(m_1)_{n_1}(m_2)_{n_2}}
\mathbb{P}_m(c,d) 
$$
Let us normalise $x=\frac{c}{m_1}$ and $y=\frac{d}{m_2}$. Notice that $\mathbb{P}_m(x,y)$ is a function from a compact space $[0,1]^8$, so by the Prohorov theorem, it has a convergent subsequence in the limit $m\rightarrow\infty$. Pick such subsequence and denote the limiting probability measure $\mu$. We then have:
\begin{equation}
p_n(\vec{\al},\vec{\be})=\int_{x,y} \prod_{i=1}^4 x_i^{a_i}y_i^{b_i} \ud\mu(x,y)
\end{equation}
To show uniqueness, consider two different measures $\mu$ and $\mu'$ with same probability $p_{n}$. All possible $\prod_{i=1}^4 x_i^{a_i}y_i^{b_i} $ generate all polynomials which form a subalgebra of continuos functions. By Stone Weierstrass the $\mu$ and $\mu'$ must be equal for all continuos functions. Then it is well known that $\mu=\mu'$.

Use the mapping the following mapping to transform $x$ and $y$ into hermitian matrices $A$ and $B$ with trace 1 by:
\begin{equation}\label{eq:Af}
 \tr (A E_i)=x_i,\quad\forall i=1,2,3,4.
\end{equation}
Denote the image of this mapping by $\mathcal{A}$. $\mathcal{A}$ is  bounded since the space of probability distributions is bounded. Lastly there is an induced measure $\nu=\mu \circ \phi$ on $(\mathcal{A},\mathcal{A}')$.
\begin{equation}\label{eq:n3}
    p(x,y)=\int \prod_{i=1}^N \tr(AE_{x_i})\prod_{j=1}^{N'} \tr(BE_{y_j}) \, d\nu(A,A').  
\end{equation}
Move the trace outside the integrals and separate $A$ and $E$ in the tensor product.
\begin{equation}\label{eq:n4}
    p(x,x') =   \tr \Big(\int  A^{\otimes N}\otimes A'^{\otimes N'}d\nu(A,A') \bigotimes_{i=1}^N E_{x_{i}}\otimes \bigotimes_{j=1}^{N'} E_{x'_{j}}\Big) 
\end{equation}
By comparing (\ref{eq:n1}) with (\ref{eq:n4}) we conclude that:
\begin{equation}\label{eq:n5}
    \rho=\int  A^{\otimes N}\otimes A'^{\otimes N'}d\nu(A,A')
\end{equation}

\noindent The only thing left is to prove that both $A$ and $A'$ are positive semidefinite. 
\begin{lemma}\label{lemmaB}
    $\nu$-a.e. $A,A'\in \mathcal{A}$ are positive semidefinite.
\end{lemma}
\begin{Proof}
We focus on $A$ here ($A'$ can be done in a similar fashion). Take the partial trace of $\rho$ eliminating $A'^{\otimes N'}$. Notice that the partial trace of a density operator is still a density operator.
$$ \tilde{\rho}=\tr_{N'}\rho =\int  A^{\otimes N}d\nu(A,A') $$
Define a compact set where $A$ is not positive semidefinite.
$$ \mc{A}_\ep = \{ (A,A')\in \mc{A} : \min_{\|\psi\|=1} \langle \psi |A|\psi \rangle   \leq -\ep \}   $$
The compactness follows from $\mc{A}$ being bounded. Assume for the sake of contradiction $\nu(\mathcal{A}_\ep) > 0$ for some $\ep>0$. Around every point $(A_0,A_0')\in \mathcal{A}_\ep$ we can pick a minimiser $|\psi_0\rangle$ and by continuity we can find some small open ball $\mathcal{B}_0\subset \mathcal{A}$ around $A_0$ where 
$$ \langle \psi_0 |B|\psi_0 \rangle   \leq -\frac{\ep}{2},\quad \forall (B,B')\in \mc{B}_0. $$
This family of balls is an open cover of the compact set $\mc{B}_\ep$ so there exists a finite subcover and hence a ball $B_0$ with positive measure $\nu(\mc{B}_0)>0$. Define the projection operator: $\Pi_0=1-|\psi_0\rangle\langle \psi_0| $. Let $N$ be even. Apply $\Pi_0^{\otimes N}$ on $\tilde{\rho}$ and take the trace:
$$ \tr (\tilde{\rho}\Pi_0^{\otimes N} )\geq \int_{\mc{B}_0} \ud \nu \, (1+\frac{\ep}{2})^{ N} + \int_{\mathcal{A}\setminus \mc{B}_0}\ud \nu \,\tr (A\Pi_0)^{ N}\geq   \nu(\mc{B}_0)(1+\frac{\ep}{2})^{ N}  $$
This leads to a contradiction as $N\rightarrow\infty$, since the LHS $\leq 1$.
\end{Proof}


\begin{thebibliography}{9}

\bibitem{GD}
G.~Athanasopoulos and D.~Ueltschi,
``Kac--Ward solution of the 2D classical and 1D quantum Ising models,''
\emph{Annales Henri Poincar\'e},
vol.~26, pp.~2955--2978, 2025.
doi:10.1007/s00023-024-01479-2


\bibitem{Shore}
A.~Ballesteros, I.~Gutiérrez-Sagredo, V.~Mariscal, and J.~J.~Relancio,
``Quantum group deformation of the Kittel--Shore model,''
\emph{J. Phys. A: Math. Theor.} \textbf{58} (2025) 445202,
arXiv:2502.20884.
doi:10.1088/1751-8121/ae1272


\bibitem{Ryan}
J.~E.~Bj\"ornberg, H.~Rosengren, and K.~Ryan,
``Heisenberg models and Schur--Weyl duality,''
\emph{Advances in Applied Mathematics}
\textbf{151}, 102572 (2023).
doi:10.1016/j.aam.2023.102572

\bibitem{Caves}
C.~M.~Caves, C.~A.~Fuchs, and R.~Schack,
``Unknown quantum states: The quantum de Finetti representation,''
\emph{J.\ Math.\ Phys.} \textbf{43}, 4537--4559 (2002).
arXiv:quant-ph/0104088.
doi:10.1063/1.1494475

\bibitem{Fannes}
M.~Fannes, H.~Spohn, and A.~Verbeure,
``Equilibrium states for mean field models,''
\emph{Journal of Mathematical Physics}
\textbf{21}, 355--358 (1980).
doi:10.1063/1.524236



\bibitem{Heath}
Heath, D., \& Sudderth, W. (1976). De Finetti’s Theorem on Exchangeable Variables. \textit{The American Statistician}, \textit{30}(4), 188–189. https://doi.org/10.1080/00031305.1976.10479175 


\bibitem{HudsonMoody1976}
R.~L. Hudson and G.~R. Moody,
\newblock Locally normal symmetric states and an analogue of de Finetti's theorem,
\newblock {\em Zeitschrift f\"ur Wahrscheinlichkeitstheorie und Verwandte Gebiete}
\textbf{33} (1976), 343--351.

\bibitem{Stormer1969}
E.~St{\o}rmer,
\newblock Symmetric states of infinite tensor products of $C^*$-algebras,
\newblock {\em Journal of Functional Analysis} \textbf{3} (1969), 48--68.







\bibitem{Warzel}
C.~Manai and S.~Warzel,
``The Spectral Gap and Low-Energy Spectrum in Mean-Field Quantum Spin Systems,''
\emph{Forum of Mathematics, Sigma} \textbf{11}, e112 (2023).
doi:10.1017/fms.2023.111
arXiv:2302.00465

\bibitem{Bergamaschi2022}
T.~Bergamaschi,
``Improved Product-state Approximation Algorithms for Quantum Local Hamiltonians,''
arXiv preprint arXiv:2210.08680 [quant-ph] (2022).



\end{thebibliography}
\end{document}